\documentclass[preprint,amsmath,amssymb,aps]{revtex4-1}
\usepackage{graphicx}
\usepackage{dcolumn}
\usepackage{pstricks}
\usepackage{bm}
\usepackage{tikz}

\begin{document}

\title{Dynamics of a spin-boson model with structured spectral density}

\author{Arzu Kurt}
\author{Resul Eryigit}
\email{resul@ibu.edu.tr}
\affiliation{
 Department of Physics, Abant Izzet Baysal University, Bolu, Turkey 
}

\date{\today}
\begin{abstract}
We report the results of a study of the dynamics of a two-state system 
coupled to an environment with peaked spectral density. An exact analytical expression for the bath correlation function is obtained. Validity range of various approximations to the correlation function for calculating the population difference of the system are discussed as 
function of tunneling splitting, oscillator frequency, coupling constant, damping rate and the temperature of the bath. An exact expression for the population difference for a limited range of parameters, is derived. 
\end{abstract}

\pacs{Valid PACS appear here}
\maketitle

\section{Introduction}

The spin-boson model is one of the most prominent models used to study dissipative and decoherence effects in quantum mechanics~\cite{weiss,legett87}. It describes a two-state system (TSS) coupled to an infinite array of non-interacting harmonic oscillators whose effect on the system is characterized by spectral density $J(\omega)$. The spin-boson model with a power law spectral density, which is the general setting of the model, contains only the cutoff frequency of the bath as an internal energy scale and leads to scale-free-rates while the so-called structured environments provide a more non-trivial internal dynamics  which might be relevant for controlling coherence and relaxation times by engineering a properly structured environment~\cite{wilhelm04}.
Garg, Onuchic and Ambegaokar (GOA) have shown that 
the spin-boson model with Ohmic spectral density can be mapped to the problem of a TSS interacting with a harmonic oscillator which is damped by an Ohmic environment whose effective spectral function can be approximated as Lorentzian when the cutoff frequency of the Ohmic bath is much larger than the characteristic frequency of the harmonic oscillator~\cite{garg85}. The GOA model has been used to in many studies to describe several phenomena such as electron transfer reactions in various condensed phase environments. The same Hamiltonian and $J(\omega)$, also, describe experimentally relevant quantum systems, such as flux-qubit read out by a dc-SQUID~\cite{vanderWal00,tian02,chiorescu03}, atom-based cavity quantum electrodynamics~\cite{raimond02}, circuit quantum electrodynamics with superconducting systems~\cite{schuster07}, semiconducting quantum dots in nano-cavities~\cite{badolato05} and in nano-mechanical resonators~\cite{blencowe04,oconnell10}. The same model is employed in chemical physics context to study charge transfer~\cite{tanaka09,dijkstra15}, energy transfer dynamics in photosynthesis~\cite{escher11,chin13,iles16} and linear and nonlinear spectroscopies~\cite{dijkstra15}. 

Theoretical approaches used to investigate the dynamics of the coupled TSS-damped harmonic oscillator(HO) system can be, broadly, divided into two groups; on the one hand, the infinite dimensional TSS-HO Hamiltonian is approximated by a finite dimensional system which is reduced to Jaynes-Cummings or ac-Stark Hamiltonian depending on whether one is in the resonant or dispersive regime~\cite{blais04}. On the other hand, the problem could be considered as a spin-boson model with a peaked spectral density. A large number of computational techniques have been developed to investigate the dynamics of spin-boson model, such as hierarchical equations of motion, renormalization group techniques and  path integral based formalism ~\cite{tanimura90,kleff04,tanaka09,prior10,chin13,dijkstra152}. The effects of structured spectral density on the decoherence properties of a qubit have been studied with a perturbative approach in Refs.~\cite{goorden04,goorden05}. Thorwart \textit{et. al.} have used ab initio QUAPI technique to show that perturbative treatment breaks down when qubit-HO coupling is strong ($g\gg\Gamma$) and when the qubit and the oscillator frequencies are comparable~\cite{thorwart04}. Gan, Huang and Zheng have studied the dynamics of the spin-boson model by using a unitary transformation method~\cite{gan10} while Ref.~\cite{kleff04} has investigated the dephasing times for the TSS-HO system by using flow-equations renormalization method and  shown that harmonic oscillator frequency can be used to control qubit dynamics.

As the interaction between the TSS and its environment, whether a bath of non-interacting harmonic oscillator or the damped harmonic oscillator, is considered to be linear, the spectral density function $J(\omega)$ and the bath temperature completely characterizes the TSS-bath coupling. All relevant computational techniques make use of bath correlation function $G(t)$ which is the thermal average of the force auto-correlation of the bath degrees of freedoms or its two-times integrated form to account for the effect of the bath. So, obtaining manageable analytical expressions for $G(t)$ might be beneficial for theoretical as well as computational techniques used to treat the spin-boson problem.  There has been a number of reports on the analytical expressions for correlation functions of peaked spectral density~\cite{garg85,tanimura06,nesi07,vierheilig11}.
 Nesi, Grifoni and Paladino have studied the dynamics of the system for the large Q-factor oscillator at arbitrary detuning and finite temperatures and obtained analytical expressions for the TSS population~\cite{nesi07} for the weak-coupling regime. Building on the findings of Ref~\cite{nesi07}, Vierheilig, Bercioux and Grifoni have considered the dynamics of a qubit coupled to a nonlinear oscillator which is coupled to an Ohmic bath and showed that the system can be mapped to the TSS-damped harmonic oscillator model with an effective peaked spectral density~\cite{vierheilig11}. 

In the present study, our first aim is to obtain an exact analytical expression for 
the correlation function $G(t)$ of TSS-HO system which is characterized by a peaked spectral density. The TSS-HO system has several characteristic times set by the energy splitting and tunneling amplitude of the TSS, frequency and the damping of the harmonic oscillator, the temperature of the HOs environment and the coupling constant between the HO and the TSS. The complicated interplay among these rate constants make it difficult to develop a universal computational technique which is applicable for all parameter values. Using the derived $G(t)$ expression, we investigate the population dynamics of the TSS and map the validity range of various approximations as function of TSS tunneling splitting, coupling  strength between the TSS and the harmonic oscillator, damping of the oscillator and the temperature of the bath. It is found that for a range of parameters, one can use Markov approximation and obtain an exact expression for the population difference. 

The outline of the paper is follows: In Section~\ref{model}, we describe the problem and derive the expression for $G(t)$ and present several approximations to it. The validity range of the approximations are discussed in Section~\ref{discuss} and a brief summary of the study is presented in the conclusions.
\section{The model\label{model}}
Let the total Hamiltonian of the closed system be 
\begin{equation}
H_{t}=
\frac{\epsilon_0}{2}\sigma_{z}+
\frac{V}{2}\sigma_{x}+
\sum_{k}\omega_{k}b^{\dagger}_{k}b_{k}+
\sigma_z\sum_{k} g_{k}\left(b^{\dagger}_{k}+b_{k}\right)
\label{eq:Hamiltonian}
\end{equation}
\noindent where $\epsilon_0$ is the splitting of the energy levels of the system, $V$ is the tunneling matrix element and $\sigma_{i}$ are the Pauli spin matrices. The bath is modeled as a collection of harmonic oscillators with creation (annihilation) operators $b^{\dagger}_k$ ($b_k$) and the mode frequency $\omega_k$. The dynamics of the two-state system are characterized by the population difference $P(t)$ which is defined as the expectation value of $\sigma_z$  as $P(t)=\mathrm{Tr}_{\mathrm{TLS}}\left[\sigma_{z}\mathrm{Tr}_{\mathrm{B}}\left(e^{-iHt}\rho(0)e^{iHt}\right)\right]$. Here, $\rho(0)$ is the initial density matrix of the total system which is assumed to be in factorized form $\rho(0)=\rho_{\mathrm{TLS}}(0)\otimes \mathrm{exp}\left(-\beta H_{\mathrm{B}}\right)/Z$ and the two partial trace operations refer to 
trace over the bath (B) and system (TLS) degrees of freedom. $P(t)$ for the spin-boson problem obeys the generalized master equation~\cite{weiss}:
\begin{equation}
\frac{dP(t)}{dt}=-\int_{0}^{t}\,dt'\left(K^{s}(t,t')P(t')+K^{a}(t,t')\right)
\label{eq:dpdt}
\end{equation}
\noindent where $K^{a}(t,t')$ and $K^{s}(t,t')$ are the asymmetric and symmetric parts of the kernel, respectively. They are derived from the two-time integrated bath correlation function $G(t)$
\begin{equation}
G(t)=\int_{0}^{t}dt_{1}\;\int_{0}^{t_{1}}\;dt_2\,\left\langle F(t_2)F(0)\right\rangle_{T}+i E_r t
\label{eq:gt}
\end{equation}
\noindent and can be 
represented as a series in tunneling amplitude $V$. In Eq.~\ref{eq:gt}, $E_r$ is the reorganization energy of the bath and  $\left\langle F(t)F(0)\right\rangle_{T}$ is the thermal average of the force auto-correlation function of the bath modes which is defined as:
\begin{equation}
\left\langle F(t)F(0)\right\rangle_{T}=\frac{1}{2\pi}\int_{0}^{\infty}J(\omega)\frac{\cosh{\left(\omega/2k_{B}T-i\omega t\right)}}{\sinh{\left(\omega/2k_{B}T\right)}}\;d\omega
\end{equation}
The effect of the environment on the system is characterized by bath spectral function $J(\omega)$, which we assume to be of the form
\begin{equation}
J(\omega)=8 \kappa^2\frac{\gamma \,\omega_0\,\omega}{\left(\omega^2-\omega_0^2\right)^2+4 \gamma^2\omega^2}
\label{eq:spectral}
\end{equation}
\noindent for the present study. $J(\omega)$ in Eq.~\ref{eq:spectral} is a structured spectral function and can be used to describe various types of environments, such as the traditional spin-boson model with an Ohmic environment~\cite{garg85}, a TSS coupled to a nonlinear, damped-oscillator~\cite{vierheilig11}, or an environment that contains both a background vibrations and one prominent vibrational mode.  Depending on the model it describes, its parameters ($\omega_0,\,\kappa,\,$ and $\gamma$) would have slightly different meanings. Here, we have a two-state system in contact with a harmonic oscillator which damped by an Ohmic thermal bath in mind. So, $\omega_0$ is the frequency of the central harmonic oscillator, $\kappa$ is the TSS-HO coupling constant, $\gamma$ is the broadening of the oscillator levels due to its interaction with the Ohmic environment. One should note that $J(\omega)$ of Eq.~\ref{eq:spectral} reduces to so called Debye form in the over-damped limit $\omega_0\ll\gamma/2$. 

The reorganization energy $E_r$ is defined as 
\begin{equation}
E_r=\frac{1}{2 \pi}\int_{0}^{\infty}\frac{J(\omega)}{\omega}\; d\omega
\end{equation}
\noindent and is equal to $\kappa^2/\omega_0$ for the spectral density given in equation~\ref{eq:spectral}. The bath correlation function $G(t)$ can be simplified for the strong coupling regime $\kappa\gg\omega_0$, where one can invoke the short-time approximation by noting that the kernel function $G(t)$ is non-zero only for a very short time and obtain~\cite{garg85}:
\begin{equation}
G_{\mathrm{st}}(t)= \left\langle F(0)^2\right\rangle t^2+iE_rt
\label{eq:gst}
\end{equation}
\noindent where $\left\langle F(0)^2\right\rangle$ can be evaluated from Eq.~\ref{eq:gt} by using Eq.~\ref{eq:spectral} as
\begin{equation}
\left\langle F(0)^2\right\rangle=2\frac{E_r}{\beta}+\frac{1}{\pi}\frac{E_r\omega_0^2}{\Delta}\,
\mathrm{Im}\left[
\psi\left(1+i\tilde{\beta}(\Delta-i\gamma)\right)\right]
\end{equation}
\noindent where $\Delta=\sqrt{\omega_0^2-\gamma^2}$, $\tilde{\beta}=\beta/(2\pi)=1/(2\pi k_B T)$ is the inverse temperature scaled by $1/(2\pi)$ and  $\psi(z)$ is the complex di-gamma function. To evaluate $G(t)$ for the general case, we write the force auto-correlation function as:
\begin{equation}
\left\langle F(t)F(0)\right\rangle_{T}=\frac{1}{2\pi}\int_{0}^{\infty}J(\omega)\left(\coth{\left(\beta\omega/2\right)}\cos{(\omega \,t)}-i\sin{(\omega\,t)}\right)\;d\omega
\label{eq:corr}
\end{equation}

\noindent One should note that the auto-correlation function is the difference of the  Fourier cosine and sine transforms of $J(\omega)\coth{\left(\beta\omega/2\right)}$ and $J(\omega)$. Fourier sine transform of the spectral function is straightforward, while the cosine transform part can be accomplished by using the series form of the hyperbolic cotangent function:
\begin{equation}
\coth{(\beta\omega/2)}=\frac{2}{\beta\omega}+\frac{4}{\beta}\sum_{m=1}^{\infty}
\frac{\omega}{\omega^2+4 m^2\pi^2/\beta^2}
\label{eq:cothapp}
\end{equation}
\noindent Using Eqs.~\ref{eq:spectral} and \ref{eq:cothapp} in Eq.~\ref{eq:corr} and carrying out the double integral in Eq.~\ref{eq:gt}, it is possible to obtain an exact expression for $G(t)$ as: 
\begin{equation}
G_{\mathrm{all}}(t)=-(a+b)+a \exp{\left(-it\,z^{\star}\right)}+b\exp{\left(-i t\,z\right)}-c\,t+\sum_{n=1}^3K_{n}
\label{eq:all}
\end{equation}
\noindent where $z=\Delta+i\gamma$, $\Delta=\sqrt{\omega_0^2-\gamma^2}$  and 
\begin{eqnarray*}
a=\frac{E_r}{\Delta}\,\frac{e^{-2i\theta}}{e^{\beta z}-1},\qquad 
b=\frac{E_r}{\Delta}\,\frac{e^{2i\theta}}{1-e^{-\beta z^{\star}}},\qquad
c=4E_r\frac{\gamma}{\beta\omega_0^2}
\end{eqnarray*}

\noindent where $\theta=\arctan{(\gamma/\Delta)}$. The last three terms of Eq.~\ref{eq:all}, which are important for the low temperature regime, are:
\begin{eqnarray}
K_1&=&-\frac{E_r}{\pi\Delta}\mathrm{Im}\left[e^{-2i\theta}\left(H(-i\tilde{\beta}z)+H(i\tilde{\beta} z)\right)\right]\\ \label{eqn:fourthTerm}
K_2&=&4E_{r}\frac{\gamma}{\pi\omega_0^2}
\log{\left(1-e^{-\frac{t}{\tilde{\beta}}}\right)}\\ \label{eqn:fifthTerm}
K_3&=&-\frac{E_r}{\pi\Delta}\mathrm{Im}\left[e^{-2i\theta}\left(B(t,-i\tilde{\beta}z)+B(t,i\tilde{\beta}z)\right)\right] \label{eqn:sixthTerm}
\end{eqnarray}
\noindent where $\tilde{\beta}=\beta/(2\pi)$, $H(z)$ is the complex Harmonic number, $B(t,z)=e^{tz}\mathcal{B}\left(e^{-\frac{2\pi}{\beta}t},1+z\right)$ and $\mathcal{B}$ denotes 
incomplete Beta function. One should note that all three $K_n$s are real.

The correlation function at $T=0$ can not be obtained from the $T=0$ limit of Eq.~\ref{eq:all} because of the form chosen for the $\coth(x)$ in Eq.~\ref{eq:cothapp}. Instead, we use the identity $\coth{(\infty)}=1$ in Eq.~\ref{eq:corr} and carry out the Fourier sine and cosine transforms to get:

%-------------- New G(t)--------
\begin{eqnarray}
\label{eq:GeneralG}
G_{\mathrm{T=0}}(t)&=&\frac{2\,\gamma\,E_{r}}{\pi^{2}\,\Omega^{2}}\,\left(\gamma_{E} +\log(\Omega\,t)\right)+\frac{E_{r}}{\pi\,\Delta\,\Omega^{2}}\left(\left(\frac{1}{2}-\frac{\theta}{\pi}\right)\left(\Delta^{2}-\gamma^{2}\right)+i\,\gamma\,\Delta\right)\nonumber\\
&&-\frac{E_{r}}{2\,\pi\,\Delta}\left(e^{2\,i\,\theta}\,\cos\left(-z^{*}\,t\right)-i\,e^{-2\,i\,\theta}\sin\left(z\,t\right)\right)\nonumber\\
&&-\frac{E_{r}}{2\,\pi^{2}\,\Delta}\,\mathrm{Im}\left[e^{2\,i\,\theta}\left(e^{-i\,z^{*}\,t}\mathrm{Ei}\left(i\,z^{*}\,t\right)+e^{i\,z^{*}\,t}\,\mathrm{Ei}\left(-i\,z^{*}\,t\right)\right)\right],
\end{eqnarray}
%------------\Gamma equals to Omega
\noindent where $\gamma_{\mathrm{E}}$ is the Euler gamma constant and $\mathrm{Ei}(z)$ is the complex exponential integral. The $\gamma=\Omega$ special case of $G(t)$ at zero temperature can not be derived in a similarly as:
\begin{eqnarray}
\label{eq:GammaEqOmega}
G_{\mathrm{sp}}(t)=&=&-\frac{E_r}{\pi\Omega} \left((2-\Omega\,t)e^{\Omega\,t}\,\mathrm{Ei}(-\Omega\,t)+(\Omega\,t+2)e^{-\Omega\,t}(\mathrm{Ei}(\Omega\,t)+i\,\pi)\nonumber\right.\\
&&\left.-4\left(\log(\Omega\,t )+\frac{i\,\pi}{2}+\gamma_{E}
   \right)\right)
\end{eqnarray}

%-----------------
\section{Discussion~\label{discuss}}
At the zero bias-case ($\epsilon_0=0$ in Eq.~\ref{eq:Hamiltonian}), the dynamics of $P(t)$ simplifies to
\begin{equation}
\frac{dP(t)}{dt}=-\int_0^t\,K(t-t')P(t')\,dt'
\label{eq:pdt}
\end{equation}
Although the relation between the bath correlation function $G(t)$ and the kernels $K(t,t')$ in Eq.~\ref{eq:pdt} is, generally, complicated, it can be simplified in certain parameter regimes. For instance, non-interacting blip approximation (NIBA) allows one to truncate the exact kernels to first order in the square of tunneling splittings ($V^2$) and obtain
\begin{equation}
K(t)=V^2\,\mathrm{e}^{-G_s(t)}\cos{\left(G_a(t)\right)}=\mathrm{Re}\left[V^2\exp{\left(-G(t)\right)}\right]
\end{equation}
NIBA is exact at zero damping, otherwise it works best for zero bias and/or large damping and/or high temperature~\cite{legett87}. NIBA form of the population dynamics can, also, be derived for the polaron transformed spin-boson problem~\cite{orth13,wilhelm04}. NIBA is, generally, justified whenever the bath relaxation time is fast compared to the time-scales of the TSS, such that bath correlation function averages out in a short time which leads to 
weak damping of the system in longer times. Although NIBA can not be justified for all the 
range of $V$, $\omega_0$, $\kappa$, $\gamma$ and $\beta$ values considered below, we will use 
its kernel form as proxy to map the validity range of approximate $G(t)$s.

The standard way to solve the integro-differential equation~\ref{eq:pdt} is to exploit its  convolution form and use Laplace transform method to express it as:
\begin{equation}
P(s)=\frac{1}{s+K(s)}
\end{equation}
\noindent where the $K(s)$ is the Laplace transform of the kernel function. For the remainder of the paper, we will try to determine the validity range of various approximations to $G_{\mathrm{all}}(t)$ as function of $\omega_0$, $\gamma$, $\kappa$, and inverse temperature $\beta$. 

The first approximation we consider is to neglect the $K_{n}$ terms in $G_{\mathrm{all}}(t)$ of Eq.~\ref{eq:all}. Using only the first three terms of $G_{\mathrm{all}}(t)$ and expressing the exponential of exponential term that contains $z^{\star}$ in series form, one gets the dynamical 
equation:
\begin{equation}
\frac{dP(t)}{dt}=-\mathrm{Re}\left[\mathrm{e}^{-(a+b)}\,\sum_{n=0}^{\infty}\frac{a^n}{n!} \int_{0}^{t}dt'\,\exp{\left(b\mathrm{e}^{-(t-t')z}-(nz^{\star}+c)(t-t')\right)}P(t')\right]
\label{eq:firstthree}
\end{equation}
\noindent It is interesting to note that Eq.~\ref{eq:firstthree} indicates that the rate of change of the population difference show similarities to sampling from a type of Gumbel extreme value distribution. Such a distribution indicates dissipation of information on the log-linear scale while the observation variable is exponential~\cite{frank14}. At low temperatures ($k_B T< V$), $a$ is small and the sum can be approximated as the first term ($n=0$). One needs to include more and more terms of the sum as the temperature increases. The Laplace transform of Eq.~\ref{eq:firstthree} can be evaluated exactly in the form of infinite series sum as:
\begin{equation}
P_{\mathrm{F3}}(s)=\frac{1}{s+K_{\mathrm{F3}}(s)}
\end{equation}
\noindent where
\begin{equation}
K_{\mathrm{F3}}(s)=\frac{1}{z}\mathrm{e}^{-(a+b)}\sum_{n=0}^{\infty}\frac{a^n}{n!}\,(-b)^{\frac{1}{z}\left(nz^{\star}+c+s\right)}\,\Gamma\left[\frac{1}{z}\left(nz^{\star}+c+s\right),-b\right]
\label{eq:kf3}
\end{equation}
\noindent where $\Gamma[z,b]$ is the lower incomplete Gamma function. As shown below, for a large range of parameters, $a$ is negligible and including only the $n=0$ term in definition of $K_{\mathrm{F3}}(s)$ is adequate. Such solutions will be denoted by subscript F3b below. Similarly, in the short-time approximation, Eq.~\ref{eq:pdt} can be solved by using Laplace transform method as 
\begin{equation}
P_{\mathrm{st}}(s)=\frac{2\sqrt{a}}{2\sqrt{a}s+w\left(\frac{E_r+is}{2\sqrt{a}}\right)}
\label{eq:st}
\end{equation}
\noindent where $w(z)$ is the plasma dispersion function. Although $P(s)$ of Eqs.~\ref{eq:kf3} and \ref{eq:st} have no inverse Laplace transform which can be expressed in terms of known elementary or special functions, its singularities and branch properties can be used to deduce the characteristics of the dynamics of the TLS for the given parameter values.

Although the dynamical Eq~\ref{eq:pdt} has no known exact analytical solution for the peak spectral density (Eq.~\ref{eq:spectral}) for arbitrary parameters, such as the frequency $\omega_0$ and the dispersion $\gamma$ of the central oscillator, the system-oscillator coupling $\kappa$ and the bath inverse temperature $\beta$, it is possible to obtain approximate analytical solutions for certain parameter regimes. For instance, one can derive an exact analytical expression for $P(t)$ in  the strong coupling regime $\kappa\gg\omega_0$, under "Markovian" regime by writing 
Eq.~\ref{eq:pdt} as 
\begin{equation}
\frac{dP(t)}{dt}=-K(t)P(t)
\label{eq:markov}
\end{equation}
\noindent where 
\begin{equation}
K(t)=\int_0^{t}K(t')dt'=\frac{1}{2}\sqrt{\frac{\pi}{a}}\mathrm{e}^{-\xi^2}
\mathrm{Im}\left[
\mathrm{erfi}(z)
\right]
\label{eq:kernelMarkov}
\end{equation}
\noindent where $z=\xi-i\sqrt{a}t$ and $\mathrm{erfi}(z)$ is the complex error function. Eq.~\ref{eq:markov} can be integrated with $K(t)$ defined as equation~\ref{eq:kernelMarkov} to obtain:
\begin{equation}
P(t)=P(0) \exp{\left\{\frac{V^2}{2a}\left[1-2\xi\mathfrak{D}(\xi)-\mathrm{e}^{-\xi^2}\left(\cos{(E_{r}t)}-\sqrt{\pi}\,\mathrm{Re}\left[z \,\mathrm{erfi}(z)\right]
\right)\right]\right\}}
\label{eq:solmarkov}
\end{equation}
\noindent where $\mathfrak{D}(\xi)$ is the Dawson integral.

\begin{figure}[!ht]
\centering
\includegraphics[width=16cm]{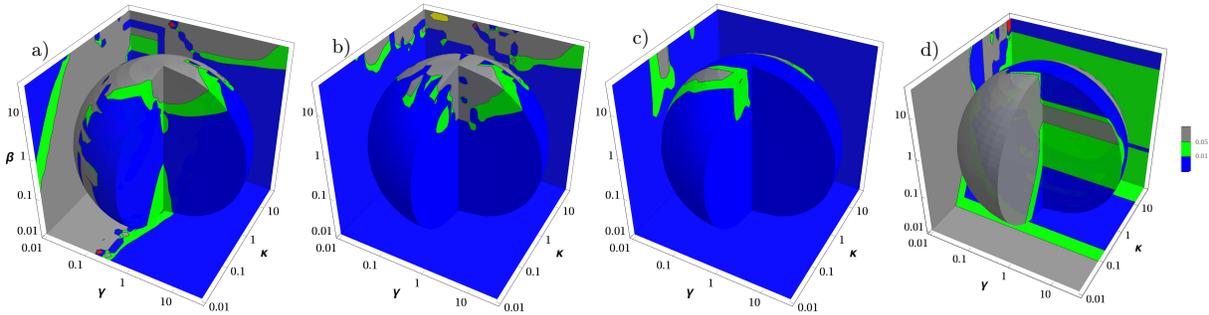}
\caption{Relative error contours of F3b (a), F3 (b), short-time (c) and Markov (d) approximations for the correlation function $G(t)$ as function 
of $\gamma$, $\kappa$ and $\beta$ for adiabatic ($\omega_0=0.1\,V$) regime.}
\label{fig:w001}
\end{figure}

As all the expressions we have found for $G(t)$ are rather involved for symbolic comparisons, 
we will use a heuristic approach to delineate the validity of approximations for various parameter regimes as follows: We solve Eq.~\ref{eq:pdt} for $P(t)$ with $G_{\mathrm{full}}(t)$ and its approximate  forms $G_{\mathrm{F3b}}(t)$, $G_{\mathrm{F3}}(t)$ and $G_{\mathrm{st}}(t)$ for a time interval $[0,t_{\mathrm{f}}]$ where $t_{\mathrm{f}}$ depends on the problem parameters $\omega_0,\,\gamma,\,\kappa,\,$ and $\beta$. $t_{\mathrm{f}}$ is chosen in such a way that  $P_{\mathrm{full}}(t)$ is almost constant at the last 1/10th of the interval. The Markovian approximation solution $P_{\mathrm{M}}(t)$ is evaluated from Eq.~\ref{eq:markov} directly. The relative error in the norm $\epsilon=\left|P_{\mathrm{app}}(t)-P_{\mathrm{full}}(t)\right|/\left|P_{\mathrm{full}}(t)\right|$ which is calculated at 1000 points of the interval $[0,t_{\mathrm{f}}]$ is used as the criteria to judge the validity of the approximation. Visual comparison suggests that for $\epsilon < 0.01$, there is no discernible difference between the two solutions. 

\begin{figure}[!ht]
\centering
\includegraphics[width=15cm]{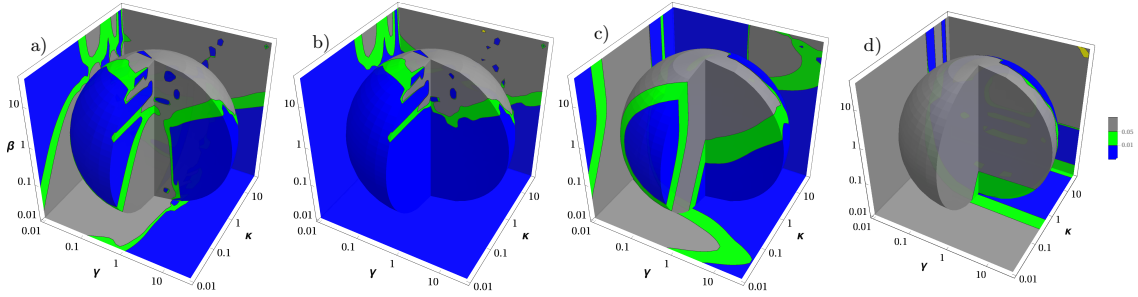}
\caption{Same as Fig.~\ref{fig:w001} for the harmonic oscillator frequency equal to the tunneling rate of the system ($\omega_0=1$).}
\label{fig:w01}
\end{figure}

\begin{figure}[!ht]
\centering
\includegraphics[width=15cm]{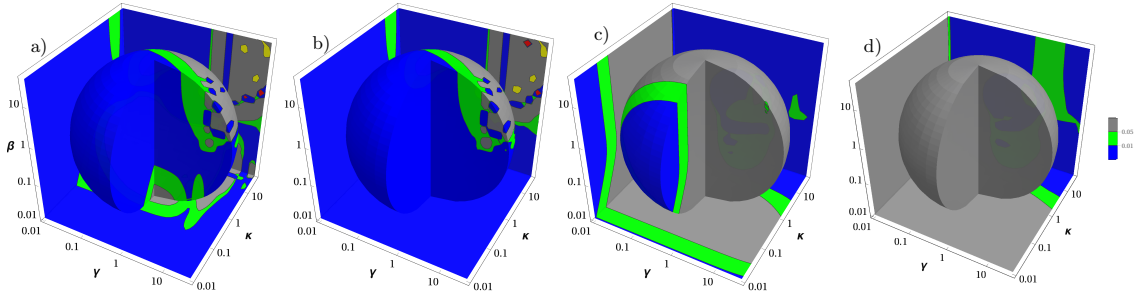}
\caption{Same as Fig.~\ref{fig:w001} for non-adiabatic regime ($\omega_0=10$).}
\label{fig:w010}
\end{figure}

We choose the time scales of the system (inverse tunneling rate) and the central oscillator ($\omega_0^{-1}$) as the base and display the relative errors for the F3b, F3, ST and M approximations in Figs.\ref{fig:w001}-~\ref{fig:w010}(a-d) for the adiabatic ($\omega_0\ll V$), intermediate ($\omega_0=V$) and the non-adiabatic ($\omega_0\gg V$) regimes, respectively. These figures display the constant relative error surfaces ($\epsilon<0.01$ (in blue), $0.01<\epsilon<0.05$ (green) and $\epsilon>0.05$ (gray)) as function of oscillator damping $\gamma$, system-oscillator interaction strength $\kappa$ and the inverse temperature of the bath $\beta$. We should emphasize that the blue regions in the figures means that the relative error $\epsilon$ is less than the arbitrarily chosen value of 0.01 for the given $\gamma$, $\kappa$ and $\beta$ and in the blue region magnitudes of $\epsilon$ for the F3b, F3, and ST approximations are not equal everywhere. One can observe from all these three figures that there are appreciable ranges of $\kappa$, $\gamma$ and $\beta$s that F3b, F3, and ST approximations produce reasonably accurate dynamics as evidenced by the appearance of the blue regions while the Markovian approximation has validity only for the strong coupling regime. 

In the adiabatic regime ($\omega_0=0.1$), the harmonic oscillator is slow compared to the dynamics of the two-state system and acts as a static disorder and the short-time approximation is found to be the approximation that has a reasonable accuracy for the widest range of $\kappa$, $\gamma$ and $\beta$ values, as can be deduced from Fig.~\ref{fig:w001}-c. Figure~\ref{fig:w01} suggests that as the oscillator response time and the characteristic time of the system become comparable, F3 approximation is justified for a large range of parameters, which means that the population dynamics is driven by a combination of Gumbel type extreme value statistics. A similar situation with a single extreme value distribution dominated dynamics is observed for the non-adiabatic regime where the oscillator response time is much faster compared to that of the system.

\begin{figure}[!ht]
\centering
\includegraphics[width=15cm]{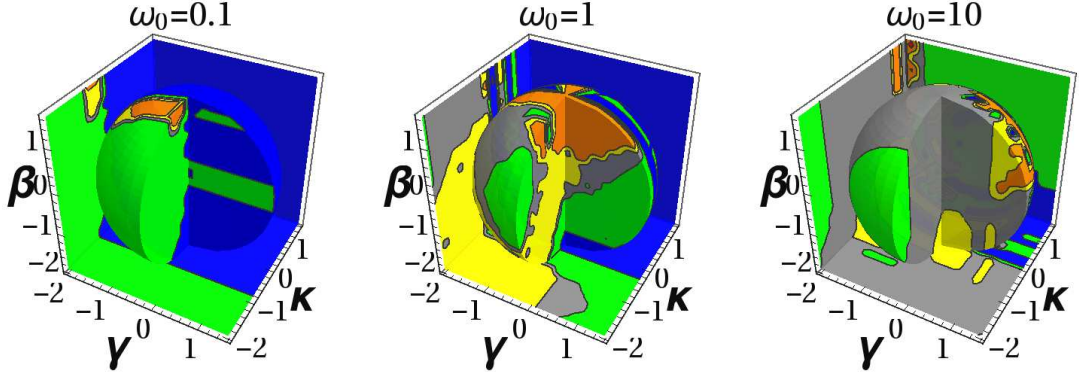}
\caption{Validity regions of various approximations for the $G(t)$ as function 
of $\gamma$, $\kappa$ and $\beta$ for adiabatic ($\omega_0=0.1\,V$), intermediate ($\omega_0=V$) and non-adiabatic ($\omega_0=10\,V$) regimes. Note that all three axis are scaled logarithmically and color coding is as blue:Markov, green:short-time, gray:F3b and yellow:F3 approximations, while the orange color indicates that one needs to include all terms in $G(t)$.}
\label{fig:w0all}
\end{figure}

Figure~\ref{fig:w0all}a-c is a summary of Figs.1-3 and displays the regions where relative error in $P(t)$ for F3b, F3, ST, and M approximations is less than 0.01 as function of inverse temperature $\beta$, coupling $\kappa$ and damping $\gamma$ for the adiabatic ($\omega_0=0.1 V$), intermediate ($\omega_0= V$) and non-adiabatic ($\omega_0=10 V$) regimes, respectively. Color-coding in the figure is such that the blue regions indicate the validity region of the Markovian solution with short time approximation, green regions show where the solution for the short-time approximation kernel has $\epsilon<0.01$, the gray and yellow colored parameter regions indicate the validity region of F3b and F3 approximations while the orange regions indicate that one needs to include all the terms in Eq.~\ref{eq:all} in the bath correlation function. One should note that in some parameter regions, more than one approximation might satisfy the inequality $\epsilon<0.01$. Since we are not searching for the best approximation at the given parameters, but the validity range of various approximations, coloring is done by checking the inequality criteria in the short-time Markovian, short-time, F3b and F3 approximations order. Changing this order leads to somewhat different coloring.

The first observation from Fig.~\ref{fig:w0all} is that the short-time approximation whether in Markovian or non-Markovian form provides an adequate description of the population dynamics for the most values of $\gamma$, $\kappa$, and $\beta$ parameters in the adiabatic case. As the adibacity is reduced, i.e. the bath response time gets faster, short-time approximation cease to describe the dynamics faithfully and gamma-type distribution becomes the valid approximation. There exist two prominent regimes in the figure; for the strong coupling between the TSS and the oscillator ( $\kappa\gg \omega_0$ ), the short-time approximation is valid for a wide temperature range at both under- and over-damped limits of $\gamma$. For the weak coupling regime, the first three terms of the kernel provides an adequate description of the dynamics.  
As the reorganization energy is increased, there is a phase transition to state where there is no tunneling. 
The full description of the kernel function is found to be necessary at low temperatures for both adiabatic and non-adiabatic regimes as indicated by the orange regions in Fig.~\ref{fig:w0all}.

\section{Conclusions}
We have studied the bath correlation function $G(t)$ for the spin-boson model with a thermal bath which is characterized by a peaked spectral density. An analytical expression which is exact for all the system parameters, such as the temperature of the bath, central frequency of the bath oscillator, system-bath coupling constant and the oscillator damping constant, is obtained. We have analyzed the population dynamics of a two-level system under various approximations to $G(t)$ in adiabatic as well as non-adiabatic conditions and found that the short-time approximation provides an accurate description of the dynamics for a large range of model parameters. Also, an exact expression for $P(t)$ is obtained under Markovian conditions 
which is found to be a valid description of the population dynamics for the strong TSS-harmonic oscillator interaction.       
\newpage
\end{document}